% Date: 1999 September (after Luis's comments)

\magnification = \magstephalf
\hsize=16.1truecm  \hoffset=0.6truecm  \vsize=23.2truecm  \voffset=-0.1truecm

\def\dblbaselines{\baselineskip=15pt \lineskip=0pt \lineskiplimit=0pt}

\def\vs{\vskip 8pt} \def\vss{\vskip 6pt} 
\parskip = 0pt % default is \parskip = 0pt+1pt
\nopagenumbers

    \font\bbf=cmbx12

\def\makeheadline{\vbox to 0pt{\vskip-30pt\line{\vbox to8.5pt{}\the
                               \headline}\vss}\nointerlineskip}
\def\toppageno{\headline={\hss\tenrm\folio\hss}}
\def\footnoterule{\kern-3pt \hrule width \hsize \kern 2.6pt \vskip 3pt}
\def\omit#1{\empty}
\pretolerance=15000  \tolerance=15000
\def\ts{\thinspace}  \def\cl{\centerline}
          
\def\ba{\kern -1pt}  \def\b{\kern -0.2em}
\def\0{\phantom{0}}    \def\00{$\phantom{000000}$}

\def\1{\phantom{1}}  
\def\etal{{\it et~al.~}} 
\def\gapprox{$_>\atop{^\sim}$}  \def\lapprox{$_<\atop{^\sim}$}
\newdimen\sa  \def\sd{\sa=.1em  \ifmmode $\rlap{.}$''$\kern -\sa$
                                \else \rlap{.}$''$\kern -\sa\fi}
              \def\se{\sa=.1em  \rlap{.}{''}\kern -\sa}
              \def\dgd{\sa=.1em \ifmmode $\rlap{.}$^\circ$\kern -\sa$
                                \else \rlap{.}$^\circ$\kern -\sa\fi}
\newdimen\sb  \def\md{\sa=.06em \ifmmode $\rlap{.}$'$\kern -\sa$
                                \else \rlap{.}$'$\kern -\sa\fi}
\def\kms{km~s$^{-1}$}
\def\s{\ifmmode ^{\prime\prime} \else $^{\prime\prime}$ \fi}
\def\min{\ifmmode ^{\prime} \else $^{\prime}$ \fi}
\def\deg{\ifmmode ^{\circ} \else $^{\circ}$ \fi}
\def\msun {M$_{\odot}$~}  \def\msund{M$_{\odot}$}  
\def\mbh{$M_{\bullet}$}   
\def\lsun{$L_\odot$~}     \def\lsund{$L_\odot$}    

% Luis Ho's TeX definitions:

\def\e#1{$\times$10$^{#1}$}

\def\pp{\parshape 2 0truein 6.5truein .3truein 6.2truein}

%  \pp is a simple definition to define a paragraph shape in 
%      which the first line is not indented, but subsequent lines are.
%      suitable for references and figure captions.

\parindent=20pt

\dblbaselines

\cl{{\bbf SUPERMASSIVE BLACK HOLES IN INACTIVE GALAXIES}\footnote{$^1$\b}
{To appear in {\it Encyclopedia of Astronomy and Astrophysics}}}

\vskip 0.5truecm

\cl{John Kormendy\footnote{$^2$\b}{Department of Astronomy, RLM 15.308,
University of Texas, Austin, TX 78712-1083}{\ts}
and Luis C. Ho\footnote{$^3$\b}{Carnegie Observatories, 813 Santa Barbara 
St., Pasadena, CA 91101-1292}}

\vskip 0.5truecm

\vs
\cl{1.~INTRODUCTION}
\vs

      Several billion years after the Big Bang, the Universe went through a 
``quasar era'' when high-energy active galactic nuclei (AGNs) were more than
10,000 times as numerous as they are now.  Quasars must then have been standard 
equipment in most large galaxies.  Since that time, AGNs have been dying out. 
Now quasars are exceedingly rare, and even medium-luminosity AGNs such as 
Seyfert galaxies are uncommon.  The only activity that still occurs in many 
galaxies is weak.  A paradigm for what powers this activity is well established
through the observations and theoretical arguments that are outlined in the 
previous article.  AGN engines are believed to be supermassive black holes 
(BHs) that accrete gas and stars and so transform gravitational potential 
energy into radiation.  Expected BH masses are \mbh\ $\sim$ 10$^6$ -- 
10$^{9.5}$ \msund.  A wide array of phenomena can be understood within this 
picture.  But the subject has had an outstanding problem: there was no dynamical
evidence that BHs exist.  The search for BHs has therefore become one of the 
hottest topics in extragalactic astronomy.

      Since most quasars have switched off, dim or dead engines -- starving 
black holes -- should be hiding in many nearby galaxies.  This means that the 
BH search need not be confined to the active galaxies that motivated it.  In
fact, definitive conclusions are much more likely if we observe objects in which
we do not, as Alan Dressler has said, ``have a searchlight in our eyes.''  Also,
it was necessary to start with the nearest galaxies, because only then could we 
see close enough to the center so that the BH dominates the dynamics.  Since 
AGNs are rare, nearby galaxies are not particularly active.  For these reasons,
it is no surprise that the search first succeeded in nearby, inactive galaxies.

      This article discusses stellar dynamical evidence for BHs in inactive and
weakly active galaxies.  Stellar motions are a particularly reliable way to 
measure masses, because stars cannot be pushed around by nongravitational 
forces.  The price is extra complication in the analysis: the dynamics are
collisionless, so random velocities can be different in different directions. 
This is impossible in a collisional gas.  As we shall see, much effort has gone
into making sure that unrecognized velocity anisotropy does not lead to 
systematic errors in mass measurements.

   Dynamical evidence for central dark objects has been published for 17
galaxies.  With the {\it Hubble Space Telescope\/} ({\it HST\/}) pursuing the 
search, the number of detections is growing rapidly.  Already we can ask 
demographic questions.  Two main results have emerged.  First, the numbers and 
masses of central dark objects are broadly consistent with predictions based on
quasar energetics.  Second, the central dark mass correlates with the mass of 
the elliptical-galaxy-like ``bulge'' component of galaxies.  What is less secure
is the conclusion that the central dark objects must be BHs and not (for 
example) dense clusters of brown dwarf stars or stellar remnants.  Rigorous 
arguments against such alternatives are available for only two galaxies.  
Nevertheless, these two objects and the evidence for dark masses at the centers
of almost all galaxies that have been observed are taken as strong evidence 
that the AGN paradigm is essentially correct.

\pageno=2 \toppageno

\vs
\cl{2.~DEAD QUASAR ENGINES IN NEARBY GALAXIES}
\vs

      The qualitative discussion of the previous section can be turned into a
quantitative estimate for \mbh\ as follows.  The quasar population produces an 
integrated comoving energy density of
$$
u = \int_0^\infty \int_0^\infty \Phi(L,z) \,L\, dL\, {dt\over{dz}\,} dz\, =\, 
1.3\times10^{-15}~{\rm erg~cm}^{-3}\ts, \eqno{(1)}
$$
where $\Phi(L,z)$ is the comoving density of quasars of luminosity $L$ at 
redshift $z$ and $t$ is cosmic time.  For a radiative energy conversion 
efficiency of $\epsilon$, the equivalent present-day mass density is $\rho_u =
u/(\epsilon c^2)  =  2.2 \times 10^4~\epsilon^{-1}$ \msun Mpc$^{-3}$.
Comparison of $\rho_u$ with the overall galaxy luminosity density, $\rho_g
\simeq 1.4 \times 10^8$\ts$h$ \lsun Mpc$^{-3}$, where the Hubble constant is
$H_{\rm 0}$ = 100\ts$h$ \kms\ Mpc$^{-1}$, implies that a typical nearby bright
galaxy (luminosity $L^* \simeq 10^{10}~h^{-2}$ \lsund) should contain a dead
quasar of mass \mbh\ $\sim$ $1.6 \times 10^6~\epsilon^{-1}~h^{-3}$ \msund.
Accretion onto a BH is expected to produce energy with an efficiency of
$\epsilon \sim 0.1$, and the best estimate of $h$ is 0.71 $\pm$ 0.06. Therefore
the typical BH should have a mass of $\sim 10^{7.7}$ \msund.  BHs in dwarf
ellipticals should have masses of $\sim 10^6$ \msund.

      In fact, the brightest quasars must have had much higher masses.  A BH 
cannot accrete arbitrarily large amounts of mass to produce arbitrarily high 
luminosities.  For a given $M_\bullet$, there is a maximum accretion rate above
which the radiation pressure from the resulting high luminosity blows away the
accreting matter.  This ``Eddington limit'' is discussed in the preceeding
article.  Eddington luminosities of $L \sim 10^{47}$ erg s$^{-1}$ $\sim$ 
10$^{14}$ \lsun require BHs of mass \mbh\ \gapprox \ts10$^9$ \msund.  These 
arguments define the parameter range of interest: \mbh\ $\sim$ 10$^6$ to 
$10^{9.5}$ \msund.  The highest-mass BHs are likely to be rare, but low-mass 
objects should be ubiquitous.  Are they?

\vs
\cl{3.~STELLAR DYNAMICAL SEARCHES FOR CENTRAL DARK OBJECTS}
\vs

      The answer appears to be ``yes''.  The majority of detections on which 
this conclusion is based are stellar-dynamical.  However, finding BHs is not
equally easy in all galaxies.  This results in important selection effects that 
need to be understood for demographic studies.  Therefore we begin with a
discussion of techniques.  We then give three examples that highlight important 
aspects of the search.  NGC 3115 is a particularly clean detection that 
illustrates the historical development of the search.  M{\ts}31 is one of the 
nearest galaxies and contains a new astrophysical phenomenon connected with BHs.
Finally, the strongest case that the central mass is a BH and not a dark cluster
of stars or stellar remnants is the one in our own Galaxy.

\vs
\cl{3.1 Stellar Dynamical Mass Measurement}
\vs

      Dynamical mass measurement is conceptually simple.  If random motions are
small, as they are in a gas, then the mass $M(r)$ within radius $r$ is $M(r) = 
V^2 r / G$.  Here $V$ is the rotation velocity and $G$ is the gravitational 
constant.  In stellar systems, some dynamical support comes from random 
motions, so $M(r)$ depends also on the velocity dispersion $\sigma$.  The 
measurement technique is best described in the idealized case of spherical 
symmetry and a velocity ellipsoid that points at the center.  Then
the first velocity moment of the \hbox{collisionless Boltzmann equation gives} 
$$ M(r) = {{V^2r}\over G} + {{\sigma_r^2r}\over 
G}~\biggl[- \ts{{d\ln{\nu}}\over{d\ln{r}}} 
- {{d\ln{\sigma_r^2}}\over{d\ln{r}}} -
\biggl(1 - {\sigma_{\theta}^2 \over \sigma_r^2}\biggr) -
\biggl(1 - {\sigma_{\phi}^2 \over \sigma_r^2}\biggr)\biggr]\ts. \eqno{(2)}$$
Here $\sigma_r$, $\sigma_{\theta}$, and $\sigma_{\phi}$ are the radial and 
azimuthal components of the velocity dispersion.  The density $\nu$ is not the
total mass density $\rho$; it is the density of the luminous tracer population
whose kinematics we measure.  We never see $\rho$, because the stars that 
contribute most of the light contribute almost none of the mass.  Therefore we
assume that $\nu(r) \propto$ volume brightness.  All quantities in Equation 2 
are unprojected.  We observe brightnesses and velocities after projection and 
blurring by a point-spread function (PSF).  Information is lost in both 
processes.  Several techniques have been developed to derive unprojected 
quantities that agree with the observations after projection and PSF 
convolution.  From these, we derive the mass distribution $M(r)$ and compare it
to the light distribution $L(r)$.  If $M/L(r)$ rises rapidly as $r \rightarrow
0$, then we have found a central dark object.

      There is one tricky problem with this analysis, and it follows directly 
from Equation 2.  Rotation and random motions contribute similarly to $M(r)$, 
but the $\sigma^2 r / G$ term is multiplied by a factor that depends on the 
velocity anisotropy and that can be less than 1.  Galaxy formation can easily
produce a radial velocity dispersion $\sigma_r$ that is larger than the 
azimuthal components $\sigma_\theta$ and $\sigma_\phi$.  Then the third and 
fourth terms inside the brackets in Equation 2 are negative; they can be as 
small as $-1$ each.  In fact, they can largely cancel the first two terms, 
because the second term cannot be larger than $+1$, and the first is $\simeq 1$ 
in many galaxies.  This explains why {\it ad hoc\/} anisotropic models have 
been so successful in explaining the kinematics of giant ellipticals without 
BHs.  But how anisotropic are the galaxies?  

    Much effort has gone into finding the answer. The most powerful technique is
to construct self-consistent dynamical models in which the density distribution
is the linear combination $\rho = \Sigma N_i \rho_i$ of the density 
distributions $\rho_i$ of the individual orbits that are allowed by the 
gravitational potential.  First the potential is estimated from the light 
distribution. Orbits of various energies and angular momenta are then calculated
to construct a library of time-averaged density distributions $\rho_i$.  
Finally, orbit occupation numbers $N_i$ are derived so that the projected and
PSF-convolved model agrees with the observed kinematics.  Some authors also
maximize $\Sigma N_i \ln{N_i}$, which is analogous to an entropy.  These 
procedures allow the stellar distribution function to be as anisotropic as it 
likes in order (e.{\ts}g.)~to try to explain the observations without a BH.  In 
the end, such models show that real galaxies are not extremely anisotropic. That
is, they do not take advantage of all the degrees of freedom that the physics 
would allow.  However, this is not something that one could take for granted.  
Because the degree of anisotropy depends on galaxy luminosity, almost all BH
detections in bulges and low-luminosity ellipticals (which are nearly isotropic)
are based on stellar dynamics, and almost all BH detections in giant ellipticals
(which are more anisotropic)  are based on gas dynamics.

\vfill\eject

\vs
\cl{3.2 NGC 3115: $M_{\bullet} \simeq 10^{9.0 \pm 0.3}$ M$_\odot$}
\vs

    One of the best stellar-dynamical BH cases is the prototypical S0 galaxy 
NGC 3115 (Fig.~1).  It is especially suitable for the BH search because it is 
very symmetrical and almost exactly edge-on.  NGC 3115 provides a good 
illustration of how the BH search makes progress.  Unlike some discoveries, 
finding a supermassive BH is rarely a unique event. Rather, an initial dynamical
case for a central dark object gets stronger as observations improve.  
Eventually, the case becomes definitive.  This has happened in NGC 3115 through
the study of the central star cluster -- a tiny, dense cusp of stars like those 
expected around a BH (Figure 1). Later, still better observations may accomplish
the next step, which is to strengthen astrophysical constraints enough so that 
all plausible BH alternatives (clusters of dark stars) are eliminated.  This has
happened for our Galaxy (\S\ts3.4) but not yet for NGC 3115.  

      The kinematics of NGC 3115 show the signature of a central dark object
(Fig.~2).  The original detection was based on the blue crosses.  Already
at resolution $\sigma_* = 0\sd44$, the central kinematic gradients are steep. 
The apparent central dispersion, $\sigma \simeq 300$ km s$^{-1}$, is 
much higher than normal for a galaxy of absolute magnitude $M_B = -20.0$.  
Therefore, isotropic dynamical models imply that NGC 3115 contains a dark mass 
\mbh\ $\simeq$ $10^{9 \pm 0.3}$ \msund.  Maximally anisotropic models allow
smaller masses, \mbh\ $\sim$ 10$^8$ \msund, but isotropy is more likely given 
the rapid rotation.

      Since that time, two generations of improved observations have become 
available.  The green points in Figure 2 were obtained with the Subarcsecond
Imaging Spectrograph (SIS) and the Canada-France-Hawaii Telescope (CFHT).  This
incorporates tip-tilt optics to improve the atmospheric PSF.  The observations
with the {\it HST\/} Faint Object Spectrograph (FOS) have still higher 
resolution.  If the BH detection is correct, then the apparent rotation and 
dispersion profiles should look steeper when they are observed at higher 
resolution.  This is exactly what is observed.  If the original dynamical models are ``reobserved'' at the improved resolution, the 
ones that agree with the new data have \hbox{\mbh\ = (1 to 2) $\times$ 10$^9$ 
\msund.}  

\vfill

%\special{psfile=/d7/st/3115_decomposition/rgb2.cps hoffset=-50 voffset=350
%                                                   hscale=45   vscale=45}

%\special{psfile=/d7/st/3115_decomposition/final/bulge.cps 
%                                                   hoffset=125 voffset=350
%                                                   hscale=45   vscale=45}

%\special{psfile=/d7/st/3115_decomposition/final/disk.cps hoffset=300 voffset=350
%                                                         hscale=45   vscale=45}

%%%\special{psfile=3115hst-3panels.cps angle=-90 hoffset=-59 voffset=281
%%%                                              hscale=68   vscale=68}

\includegraphics{3115hst-3panels-small.cps}

      {\bf Figure 1.} {\it HST\/} WFPC2 images of NGC 3115.  The left panel 
shows a color image made from 1050 s $V$- and $I$-band images.  The right panel 
shows a model of the nuclear disk.  The center panel shows the difference; it 
emphasizes the compact nuclear star cluster.  Brightness is proportional to the
square root of intensity.  All panels are 11\sd6 square.  [This figure is taken
from Kormendy \etal 1996, {\it Astrophys.~J.~Lett.\/}, {\bf 459}, L57.]

\eject

     Finally, a definitive detection is provided by the {\it HST\/} observations
of the nuclear star cluster.  Its true velocity dispersion is underestimated in
Figure 2, because the projected value includes bulge light from in front of and
behind the center. When this light is subtracted, the velocity dispersion of the
nuclear cluster proves to be $\sigma = 600 \pm 37$ km s$^{-1}$.  This is the 
highest dispersion measured in any galactic center.  The velocity of escape from
the nucleus would be much smaller, $V_{\rm esc} \simeq 352$ km s$^{-1}$, if it
consisted only of stars.  Without extra mass to bind it, the cluster would fling
itself apart in $\sim 2 \times 10^4$ yr. Independent of any velocity anisotropy,
the nucleus must contain an unseen object of mass \mbh\ $\simeq 10^9$ \msund. 
This is consistent with the modeling results.  The dark object is more than 25 
times as massive as the visible star cluster.  We know of no way to make a star
cluster that is so nearly dark, especially without overenriching the visible 
stars with heavy elements.  The most plausible explanation is a BH.  This would
easily have been massive enough to power a quasar.
      
\vfill

% \cl{\null} \vskip 12.6truecm

\includegraphics{kho-figure2.cps}

      {\bf Figure 2.} Rotation velocities (lower panel) and velocity dispersions
(upper panel) along the major axis of NGC 3115 as observed at three different 
spatial resolutions.  Resolution $\sigma_*$ is the Gaussian dispersion radius of
the PSF; in the case of the {\it HST\/} observations, this is negligible
compared to the aperture size of 0\sd21.  [This figure is adapted from Kormendy
\etal 1996, {\it Astrophys.~J.~Lett.\/}, {\bf 459}, L57.]

%\vs

\eject

\vs
\cl{3.3 M{\ts}31: $M_{\bullet} \simeq 3 \times \ba10^7~\ba M_\odot$}
\vs

      M{\ts}31 is the highest-luminosity galaxy in the Local Group.  At a 
distance of 0.77 Mpc, it is the nearest giant galaxy outside our own.  It can 
therefore be studied in unusual detail.

      M{\ts}31 contains the nearest example of a nuclear star cluster embedded
in a normal bulge.  When examined with {\it HST\/}, the nucleus appears double
(Figure 3).  This is very surprising.  At a separation of $2r = 0\sd49 = 1.7$ 
pc, a relative velocity of 200 \kms~implies a circular orbit period of 50,000 
yr.  If the nucleus consisted of two star clusters in orbit around each other,
as Fig.~3 might suggest, then dynamical friction would make them merge within 
a few orbital times.  Therefore it is unlikely that the simplest possible 
explanation is correct: we are not observing the last stages of the digestion of
an accreted companion galaxy.

      The nucleus rotates rapidly and has a steep velocity dispersion gradient
(Figure 3).  Dynamical analysis shows that M{\ts}31 contains a central dark mass
\mbh\ $\simeq 3 \times \ba10^7$ \msund.  The possible effects of velocity 
anisotropy have been checked and provide no escape.  Furthermore, the asymmetry 
provides an almost independent check of the BH mass, as follows.

      The top panel of Figure 3 shows the {\it HST\/} image at the same scale as
and registered in position with the kinematics.  It shows that the dispersion
peak is approximately centered on the fainter nucleus.  In fact, it is centered
almost exactly on a cluster of blue stars that is embedded in this nucleus.
This suggests that the BH is in the blue cluster.  This hypothesis can be tested
by finding the center of mass of the asymmetric distribution of starlight plus a
dark object in the blue cluster.  The mass-to-light ratio of the stars is
provided by dynamical models of the bulge at larger radii.  If the galaxy is in 
equilibrium, then the center of mass should coincide with the center of the 
bulge.  It does, provided that \mbh\ $\simeq$ $3 \times 10^7$ \msund.  
Remarkably, the same BH mass explains the kinematics and the asymmetry of the 
nucleus.

      An explanation of the mysterious double nucleus has been proposed by
Scott Tremaine.  He suggests that both nuclei are part of a single eccentric 
disk of stars.  The brighter nucleus is farther from the barycenter; it results
from the lingering of stars near the apocenters of very elongated orbits.  The
fainter nucleus is produced by an increase in disk density toward the center.
The model depends on the presence of a BH to make the potential almost 
Keplerian; then the alignment of orbits in the eccentric disk may be maintained
by the disk's self-gravity.  Tremaine's model was developed to explain the
photometric and kinematic asymmetries as seen at resolution $\sigma_* \simeq
0\sd5$.  It is also consistent with the data in Figure 3 ($\sigma_* \simeq 
0\sd27$).  The high velocity dispersion near the BH, the low dispersion in the
offcenter nucleus, and especially the asymmetric rotation curve are signatures
of the eccentric, aligned orbits.

      Most recently, spectroscopy of M{\ts}31 has been obtained with the {\it
HST\/} Faint Object Camera.  This improves the spatial resolution by an 
additional factor of $\sim$\ts5.  At this resolution, there is a 0\sd25 wide 
region centered on the faint nucleus in which the velocity dispersion is 
\hbox{$440 \pm 70$ km s$^{-1}$.}  This is further confirmation of the existence
and location of the BH.

\vfill\eject

\cl{\null}

\vskip 14.2 truecm

\includegraphics{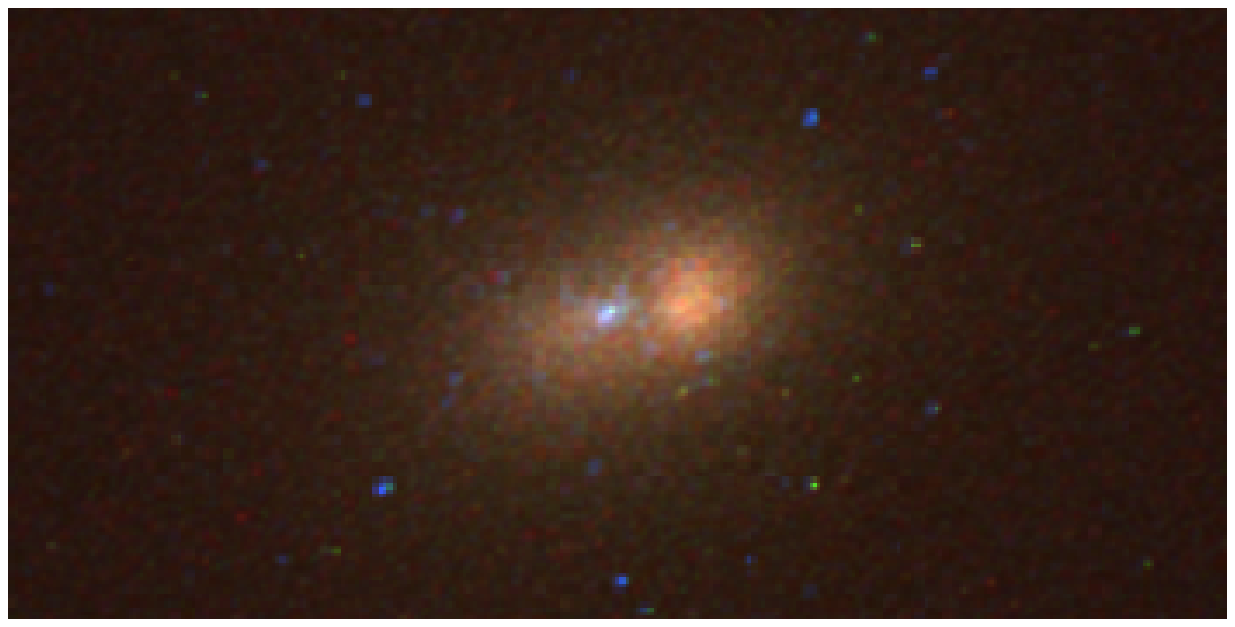}

\includegraphics{m31encyc.cps}

{\bf Figure 3.} ({\it top\/}) {\it HST\/} WFPC2 color image of M{\ts}31
constructed from $I$-, $V$- and 3000 \AA-band, PSF-deconvolved images obtained
by Lauer \etal (1998, {\it Astron.~J.\/}, {\bf 116}, 2263).  The scale is
0\sd0228 pixel$^{-1}$.  ({\it bottom and middle\/}) Rotation curve $V(r)$ and
\hbox{velocity dispersion profile $\sigma(r)$} of the nucleus with foreground 
bulge light subtracted.  The symmetry point of the rotation curve and the sharp
dispersion peak suggest that the BH is in the blue star cluster embedded in the
left brightness peak.  [This figure is adapted from Kormendy \& Bender 1999, 
{\it Astrophys.~J.\/}, {\bf 522}, 772.]

\vs

     We do not know whether the double nucleus is the cause or an effect of the 
offcenter BH.  However, offcenter BHs are an inevitable consequence of 
hierarchical structure formation and galaxy mergers.  If most large galaxies 
contain BHs, then mergers produce binary BHs and, in three-body encounters, BH 
ejections with recoil.  How much offset we see, and indeed whether we see two 
BHs or one or none at all, depend on the relative rates of mergers, dynamical 
friction, and binary orbit decay.  Offcenter BHs may have much to tell us about
these and other processes.  Already there is evidence in NGC 4486B for a second
double nucleus containing a BH.

%     These results illustrate how the early phase of the search that emphasized
%BH discovery is coming to an end; research now concentrates on the astrophysics
%of galaxies that contain BHs.  

\vs
\cl{3.4 Our Galaxy: $M_{\bullet} \simeq (2.9 \pm 0.4) \times 
                                                    \ba10^6~\ba M_\odot$}
\vs

      Our Galaxy has long been known to contain the exceedingly compact radio 
source \hbox{Sgr A*.}  Interferometry gives its diameter as 63\ts$r_s$ by less
than 17\ts$r_s$, where $r_s = 0.06$ AU = \hbox{$8.6 \times 10^{11}$ cm} is the
Schwarzschild radius of a $2.9 \times 10^6$ \msun BH.  It is easy to be 
impressed by the small size.  But as an AGN, Sgr A* is feeble: its radio 
luminosity is only $10^{34}$ erg s$^{-1}$ $\simeq 10^{0.4} L_\odot$.  The 
infrared and high-energy luminosities are higher, but there is no compelling 
need for a BH on energetic grounds.  To find out whether the Galaxy contains a 
BH, we need dynamical evidence.

      Getting it has not been easy.  Our Galactic disk, which we see in the sky
as the Milky Way, contains enough dust to block all but $\sim$\ts$10^{-14}$ of 
the optical light from the Galactic center.  Measurements of the region around
Sgr A* had to await the development of infrared detectors.  Much of the infrared
radiation is in turn absorbed by the Earth's atmosphere, but there is a useful
transmission window at 2.2 $\mu$m wavelength.  Here the extinction toward the
Galactic center is a factor of $\sim$\ts20.  This is large but manageable.  
Early infrared measurements showed a rotation velocity of $V \simeq 100$ km 
s$^{-1}$ and a small rise in velocity dispersion to $\sim 120$ km s$^{-1}$ at 
the center.  These were best fit with a BH of mass \mbh\ $\sim$ 10$^6$ \msund,
but the evidence was not very strong.  Since then, a series of spectacular
technical advances have made it possible to probe closer and closer to the 
center.  As a result, the strongest case for a BH in any galaxy is now our own.

% A_V: revised (supersedes value below) = 34 mag from Astrophysical Quantities
%      4th edition, page 159.
% A_V: Becklin et al. 1978, ApJ, 220, 831 agrees with Nuclei of Normal Galaxies
% A_2.2: footnote on p. 432 of Genzel et al. 1994, Rep. Prog. Phys. 57, 417
%        Also, Lacy (private comm.) says A_2.2 = 3 mag is correct to 20 %.

      Most remarkably, two independent groups led by Reinhard Genzel and Andrea
Ghez have used speckle imaging to measure proper motions -- the velocity 
components perpendicular to the line of sight -- in a cluster of stars at radii
$r$ \lapprox \ts0\sd5 $\simeq$ 0.02 pc from Sgr A* (Figure 4).  When combined
with complementary measurements at larger radii, the result is that the 
one-dimensional velocity dispersion increases smoothly to $420 \pm 60$ km 
s$^{-1}$ at $r \simeq 0.01$ pc. Stars at this radius revolve around the Galactic
center in a human lifetime!  The mass $M(r)$ inside radius $r$ is shown in 
Figure 5.  Outside a few pc, the mass distribution is dominated by stars, but
as $r \rightarrow 0$, $M(r)$ flattens to a constant, \mbh\ = $(2.9 \pm 0.4) 
\times \ba10^6~\ba M_\odot$.  Velocity anisotropy is not an uncertainty; it 
is measured directly and found to be small.  The largest dark cluster that is 
consistent with these data would have a central density of $4 \times 10^{12}$
\msun pc$^{-3}$.  This is inconsistent with astrophysical constraints (\S\ts5). 
Therefore, if the dark object is not a BH, the alternative would have to be
comparably exotic.  It is prudent to note that rigorous proof of a BH requires
that we spatially resolve relativistic velocities near the Schwarzschild radius.
This is not yet feasible.  But the case for a BH in our own Galaxy is now very 
compelling.

\vfill\eject

\cl{\null} \vfill

%%%\special{psfile=eckart5.eps angle=-90 hoffset=0 voffset=210 
%%%                                      hscale=70  vscale=70}

\includegraphics{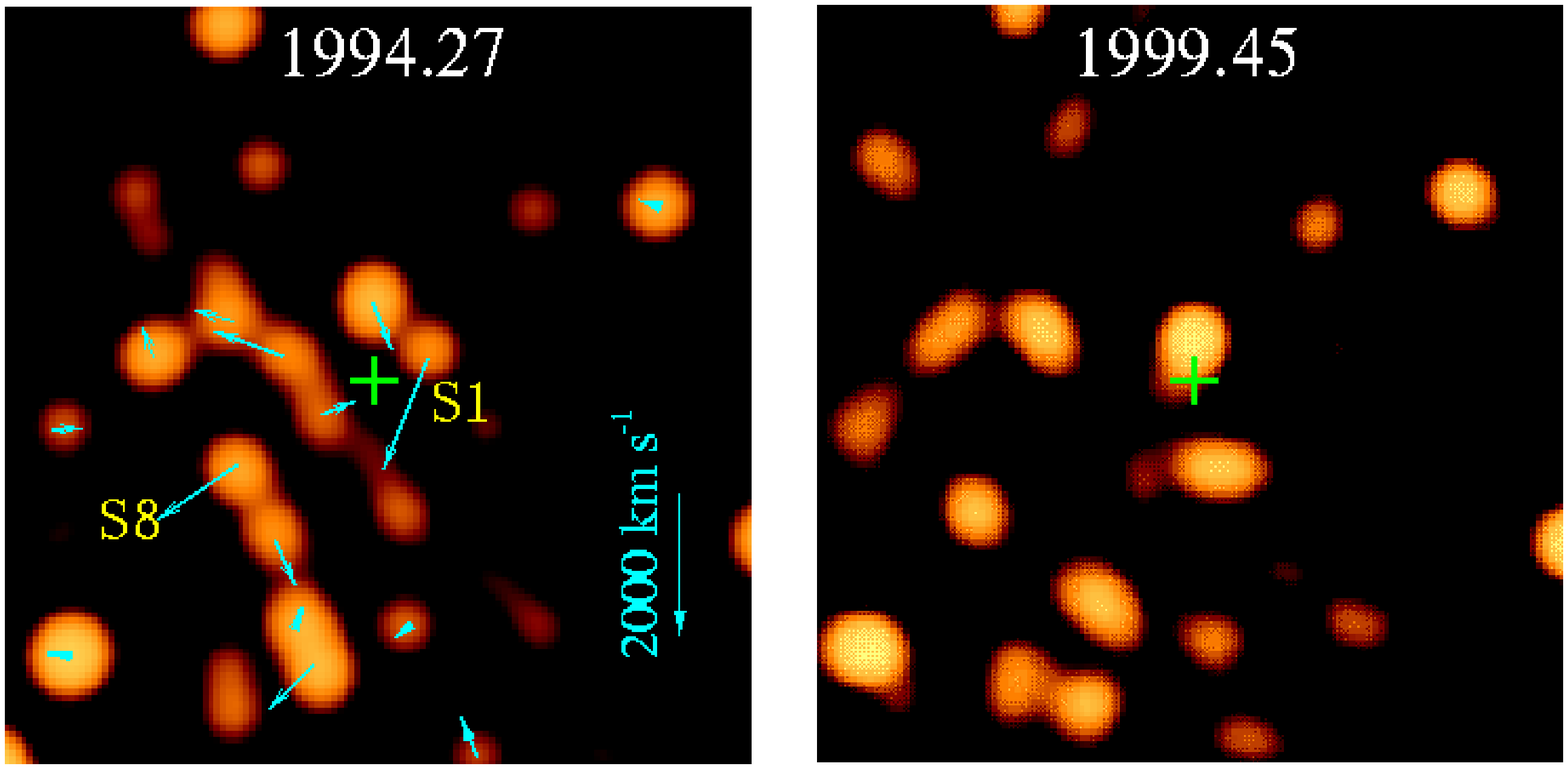}

{\bf Figure 4.} Images of the star cluster surrounding Sgr A* (green cross) at
the epochs indicated.  The arrows in the left frame show approximately where the
stars have moved in the right frame.  Star S1 has a total proper motion of 
$\sim 1600$ km s$^{-1}$.  [This figure is updated from Eckart \& Genzel 1997,
{\it M.~N.~R.~A.~S.}, {\bf 284}, 576 and was kindly provided by A.~Eckart.]

\cl{\null}\vfill

\includegraphics{gcmassaug99-ifa2.cps}

{\bf Figure 5.} Mass distribution implied by proper motion and radial velocity
measurements (blue points and curve).  Long dashes (green) show the mass 
distribution of stars if the infrared mass-to-light ratio is 2.  The red  
curve represents the stars plus a point mass $M_\bullet = 2.9 \times 10^6$ 
\msund.  Short green dashes provide an estimate of how non-pointlike the 
dark mass could be: its $\chi^2$ value is 1\ts$\sigma$ worse than the solid 
curve.  This dark cluster has a core radius of 0.0042 pc and a central density 
of $4 \times 10^{12}$ \msun pc$^{-3}$.  [This figure is updapted from Genzel 
\etal 1997, {\it M.~N.~R.~A.~S.}, {\bf 291}, 219 and was kindly provided by
R.~Genzel.]

\eject

\vs
\cl{4. BH DEMOGRAPHICS}
\vs

      The census of BH candidates as of January 2000 is given in Table 1.  The 
table is divided into three groups -- detections based on stellar dynamics, on
ionized gas dynamics, and on maser disk dynamics (top to bottom).  The rate of 
discovery is accelerating as {\it HST\/} pursues the search.  However, we
already have candidates that span the range of predicted masses and that occur 
in essentially every type of galaxy that is expected to contain a BH.  Host
galaxies include giant AGN ellipticals (the middle group), Seyfert galaxies 
(NGC 1068), normal spirals with moderately active nuclei (e.{\ts}g., NGC 4594 
and NGC 4258), galaxies with exceedingly weak nuclear activity (our Galaxy and
M{\ts}31), and completely inactive galaxies (M{\ts}32 and NGC 3115).

\vskip -7pt

\def\x{\ts\ba$\times$\null}

  \def\s{\null}  
  \def\ba{\kern -1pt}
\parskip = 0pt % default is \parskip = 0pt+1pt
\def\sup1{$^{\rm 1}$} \def\sup2{$^{\rm 2}$}
\def\r0{$\rho_0$}   
\def\00{$\phantom{000000}$} \def\0{\phantom{0}} \def\bb{\kern -2pt}

\def\1{\phantom{1}}         

%
%  VERSION DOCTORED BY JK FOR TABLES ONLY
%
%----------%
%  TABLES  %
%----------%
\def\endtable{\endgroup}
\def\tableheight{\vrule width 0pt height 8.5pt depth 3.5pt}
{\catcode`|=\active \catcode`&=\active 
    \gdef\tabledelim{\catcode`|=\active \let|=\vbar
                     \catcode`&=\active \let&=\nobar} }
\def\table{\begingroup
    \def\twidth{\hsize}
    \def\tablewidth##1{\def\twidth{##1}}
    \def\defaultheight{\vrule width 0pt height 8.5pt depth 3.5pt}
    \def\heightdepth##1{\dimen0=##1
        \ifdim\dimen0>5pt 
            \divide\dimen0 by 2 \advance\dimen0 by 2.5pt
            \dimen1=\dimen0 \advance\dimen1 by -5pt
            \vrule width 0pt height \the\dimen0  depth \the\dimen1
        \else  \divide\dimen0 by 2
            \vrule width 0pt height \the\dimen0  depth \the\dimen0 \fi}
    \def\spacing##1{\def\defaultheight{\heightdepth{##1}}}
    \def\nextheight##1{\noalign{\gdef\tableheight{\heightdepth{##1}}}}
    \def\end{\cr\noalign{\gdef\tableheight{\defaultheight}}}
    \def\zerowidth##1{\omit\hidewidth ##1 \hidewidth}    
    \def\hline{\noalign{\hrule}}
    \def\skip##1{\noalign{\vskip##1}}
    \def\bskip##1{\noalign{\hbox to \twidth{\vrule height##1 depth 0pt \hfil
        \vrule height##1 depth 0pt}}}
    \def\header##1{\noalign{\hbox to \twidth{\hfil ##1 \unskip\hfil}}}
    \def\bheader##1{\noalign{\hbox to \twidth{\vrule\hfil ##1 
        \unskip\hfil\vrule}}}
    \def\spanloop{\span\omit \advance\mscount by -1}
    \def\extend##1##2{\omit
        \mscount=##1 \multiply\mscount by 2 \advance\mscount by -1
        \loop\ifnum\mscount>1 \spanloop\repeat \ \hfil ##2 \unskip\hfil}
    \def\vbar{&\vrule&}
    \def\nobar{&&}
    \def\hdash##1{ \noalign{ \relax \gdef\tableheight{\heightdepth{0pt}}
        \toks0={} \count0=1 \count1=0 \putout##1\end 
        \toks0=\expandafter{\the\toks0 &\end} \xdef\piggy{\the\toks0} }
        \piggy}
    \let\e=\expandafter
    \def\putspace{\ifnum\count0>1 \advance\count0 by -1
        \toks0=\e\e\e{\the\e\toks0\e&\e\multispan\e{\the\count0}\hfill} 
        \fi \count0=0 }
    \def\putrule{\ifnum\count1>0 \advance\count1 by 1
        \toks0=\e\e\e{\the\e\toks0\e&\e\multispan\e{\the\count1}\leaders\hrule\hfill}
        \fi \count1=0 }
    \def\putout##1{\ifx##1\end \putspace \putrule \let\next=\relax 
        \else \let\next=\putout
            \ifx##1- \advance\count1 by 2 \putspace
            \else    \advance\count0 by 2 \putrule \fi \fi \next}   }
\def\tablespec#1{
    \def\vdimens{\noexpand\tableheight}
    \def\tabby{\tabskip=0pt plus100pt minus100pt}
    \def\r{&################\tabby&\hfil################\unskip}
    \def\c{&################\tabby&\hfil################\unskip\hfil}
    \def\l{&################\tabby&################\unskip\hfil}
    \edef\templ{\noexpand\vdimens ########\unskip  #1 
         \unskip&########\tabskip=0pt&########\cr}
    \tabledelim
    \edef\body##1{ \vbox{
        \tabskip=0pt \offinterlineskip
        \halign to \twidth {\templ ##1}}} }

\def\dot{\0\ts\raise 0.2em\hbox{{\dots}}}

$$
\table
\tablewidth{15.0truecm}
\tablespec{\l\l\c\c\c\c}
\body{
\header{{\bf Table 1} \quad Census of Black Hole Candidates}
\skip{9pt}
\hline
\skip{.2truecm}\hline
\skip{4pt}
& Galaxy & Type & $D$\0   & $M_{B,{\rm bulge}}$ & $M_\bullet$ & log{\ts}${M_\bullet}\over{M_{\rm bulge}}$ & \end
&        &      & (Mpc)\0 &                     & (\msund)    &                                           & \end
\skip{4pt}
\hline
\skip{4pt}
& Galaxy   &\0Sbc&{\0\0}0.0085    &$ -17.65 $&3\x$10^6$ &$-3.62$\0& \end
& M{\ts}31 &\0Sb &{\0\0}0.7\0\0\0 &$ -18.82 $&3\x$10^7$ &$-3.31$\0& \end
& M{\ts}32 &\0E  &{\0\0}0.7\0\0\0 &$ -15.51 $&3\x$10^6$ &$-2.27$\0& \end
& NGC 3115 &\0S0/&{\0\0}8.4\0\0\0 &$ -19.90 $&1\x$10^9$ &$-1.92$\0& \end
& NGC 4594 &\0Sa/&{\0\0}9.2\0\0\0 &$ -21.21 $&1\x$10^9$ &$-2.69$\0& \end
& NGC 3377 &\0E  &{\0\0}9.9\0\0\0 &$ -18.80 $&8\x$10^7$ &$-2.24$\0& \end
& NGC 3379 &\0E  &{\0\0}9.9\0\0\0 &$ -19.79 $&1\x$10^8$ &$-2.96$\0& \end
& NGC 4342 &\0S0 & {\0}15.3\0\0\0 &$ -17.04 $&3\x$10^8$ &$-1.64$\0& \end
& NGC 4486B&\0E  & {\0}15.3\0\0\0 &$ -16.66 $&6\x$10^8$ &$-1.03$\0& \end
\skip{4pt}
\hline
\skip{4pt}
& M{\ts}87 &\0E  &{\0}15.3\0\0\0  &$ -21.42 $&3\x$10^9$  &$-2.32$\0& \end
& NGC 4374 &\0E  &{\0}15.3\0\0\0  &$ -20.96 $&1\x$10^9$  &$-2.53$\0& \end
& NGC 4261 &\0E  &{\0}29.\0\0\0\0 &$ -20.89 $&5\x$10^8$  &$-2.92$\0& \end
& NGC 7052 &\0E  &{\0}59.\0\0\0\0 &$ -21.31 $&3\x$10^8$  &$-3.31$\0& \end
& NGC 6251 &\0E  &   106.\0\0\0\0 &$ -21.81 $&6\x$10^8$  &$-3.18$\0& \end
\skip{4pt}
\hline
\skip{4pt}
& NGC 4945 &\0Scd/&{\0\0}3.7\0\0\0&$ -15.1\0$&1\x$10^6$&$ \dot  $\0& \end
& NGC 4258 &\0Sbc&{\0\0}7.5\0\0\0 &$ -17.3\0$&4\x$10^7$&$ -2.05 $\0& \end
& NGC 1068 &\0Sb & {\0}15.\0\0\0\0&$ -18.8\0$&1\x$10^7$&$ \dot  $\0& \end
\skip{4pt}
\hline
}
\endtable
$$
Notes to Table 1: 
Column 1: galaxy name; 
column 2: Hubble type; / means that the galaxy is edge-on; 
column 3: distance based on a Hubble constant of 80 km s$^{-1}$ Mpc$^{-1}$; 
column 4: absolute $B$-band magnitude of the bulge component of the galaxy; 
column 5: BH mass based on isotropic models; 
column 6: ratio of BH mass to bulge mass.  The mass in stars is calculated from
          the luminosity via the mass-to-light ratio measured at large radii.  

\vs

      However, no complete sample has been studied at high resolution.~The 
detections in \hbox{Table 1,} together with low-resolution studies of larger 
samples of galaxies, support the hypothesis that BHs live in virtually every 
galaxy with a substantial bulge component.  The total mass in detected remnants
is consistent with predictions based on AGN energetics, within the rather large
estimated errors in both quantities.

      The main new demographic result is an apparent correlation between BH mass
and the luminosity of the bulge part of the galaxy.  This is shown in Figure 6. 
Note that the correlation is not with the total luminosity: if the disk is 
included, the correlation is considerably worse. Whether the correlation is real
or not is still being tested.  The concern is selection effects.  High-mass BHs 
in small galaxies are easy to see, so their scarcity is real.  But low-mass BHs
can hide in giant galaxies, so the correlation may be only the upper envelope of
a distribution that extends to smaller \mbh. If it is real, then the correlation
implies that BH formation or feeding is connected with the mass of the 
high-density, elliptical-galaxy-like part of the galaxy.  With the possible
exception of NGC 4945 (a late-type galaxy for which the existence and luminosity
of a bulge are uncertain), BHs have been found only in the presence of a bulge.
However, the limits on \mbh\ in bulgeless galaxies like M{\ts}33 are still
consistent with the correlation.  Current searches concentrate on the question
of whether small BHs -- ones that are significantly below the apparent 
correlation -- can be found or excluded.

      BH mass fractions are listed in Table 1 for cases in which the 
mass-to-light ratio of the stars has been measured.  The median BH mass fraction
is 0.29\ts\%.  The quartiles are 0.07\ts\% and 0.9\ts\%.

\vfill

\includegraphics{latest-encyc.cps}

{\bf Figure 6.} Correlation of BH mass with the absolute magnitude of the
bulge component of the host galaxy.  Since $M/L$ varies little from bulge to
bulge, this implies a correlation between BH mass and bulge mass.  Blue filled 
circles indicate \mbh\ measurements based on stellar dynamics, green diamonds 
are based on ionized gas dynamics, and red squares are based on maser disk 
dynamics.  It is reassuring that all three techniques are consistent with the 
same correlation.

\eject

\vs
\cl{5. ARE THEY REALLY BLACK HOLES?}
\vs

      The discovery of dark objects with masses \mbh\ $\simeq$ 10$^6$ to 
10$^{9.5}$ \msun in galactic nuclei is secure. But are they BHs? Proof requires
measurement of relativistic velocities near the Schwarzschild radius, $r_s
\simeq 2{\ts}M_\bullet/(10^8~M_\odot)$ AU.  Even for M{\ts}31, $r_s \sim 8
\times 10^{-7}$ arcsec.  {\it HST\/} spectroscopic resolution is only 0\sd1.
The conclusion that we are finding BHs is based on physical arguments that BH 
alternatives fail to explain the masses and high densities of galactic nuclei.

      The most plausible BH alternatives are clusters of dark objects produced
by ordinary stellar evolution.  These come in two varieties, failed stars and 
dead stars.  Failed stars have masses $m_*$ \lapprox \ts0.08 \msund.  They never
get hot enough for the fusion reactions that power stars, i.{\ts}e., the 
conversion of hydrogen to helium.  They have a brief phase of modest brightness
while they live off of gravitational potential energy, but after this, they 
could be used to make dark clusters.  They are called brown dwarf stars, and
they include planetary mass objects.  Alternatively, a dark cluster could be
made of stellar remnants -- white dwarfs, which have typical masses of 0.6
\msund; neutron stars, which typically have masses of $\sim$ 1.4 \msund, and
black holes with masses of several \msund.  Galactic bulges are believed to form
in violent starbursts, so massive stars that turn quickly into dark remnants
would be no surprise.  It is not clear how one could make dark clusters with the
required masses and sizes, especially not without polluting the remaining stars
with more metals than we see.  But in the absence of direct proof that the dark
objects in galactic nuclei are BHs, it is important to examine alternatives.

      However, dynamical measurements tell us more than the mass of a potential
BH.  They also constrain the maximum radius inside which the dark stuff must
live.  Its minimum density must therefore be high, and this rules out the above
BH alternatives in our Galaxy and in NGC 4258.  High-mass remnants such as white
dwarfs, neutron stars, and stellar BHs would be relatively few in number.  The
dynamical evolution of star clusters is relatively well understood; in the above
galaxies, a sparse cluster of stellar remnants would evaporate completely in 
\lapprox $10^8$ yr.  Low-mass objects such as brown dwarfs would be so numerous
that collision times would be short.  Stars generally merge when they collide.
A dark cluster of low-mass objects would become luminous because brown dwarfs 
would turn into stars. 

      More exotic BH alternatives are not ruled out by such arguments.  For
example, the dark matter that makes up galactic halos and that accounts for most
of the mass of the Universe may in part be elementary particles that are cold 
enough to cluster easily.  It is not out of the question that a cluster of 
these could explain the dark objects in galaxy centers without getting into
trouble with any astrophysical constraints.  So the BH case is not rigorously 
proved.  What makes it compelling is the combination of dynamical evidence and 
the evidence from AGN observations.  This is discussed in the previous article.

     For many years, AGN observations were decoupled from the dynamical evidence
for BHs.  This is no longer the case.  Dynamical BH detections are routine.  
The search itself is no longer the main preoccupation; we can concentrate on 
physical questions.  New technical developments such as better X-ray satellites
ensure that progress on BH astrophysics will continue to accelerate.

\vs
\cl{6.~SUGGESTIONS FOR FURTHER READING}
\vs

\frenchspacing
\parindent=8pt
\def\nhi{\noindent \hangindent=8pt}
\def\pp{\parshape 2 0truein 6.5truein .3truein 6.2truein}

%  \pp is a simple definition to define a paragraph shape in 
%      which the first line is not indented, but subsequent lines are.
%      suitable for references and figure captions.
\def\ajo{{\it Astron. J.\/}~}
\def\apjo{{\it Astrophys. J.\/}~}
\def\aplo{{\it Astrophys. J. Lett.\/}~}
\def\nato{{\it Nature\/}~}
\def\mnraso{{\it M.N.R.A.S.\/}~}
\def\annrevo{{\it Ann. Rev. Astr. Astrophys.\/}~}

\nhi$\bullet$
The search for BHs is reviewed in the following papers:

\pp
Kormendy, J., \& Richstone, D. \annrevo {\bf 33}, 581 (1995)

\pp
Richstone, D., \etal \nato {\bf 395}, A14 (1998)
\bigskip

\nhi$\bullet$
In the following papers, quasar energetics are used to predict the masses of 
dead AGN engines:

\pp
So\l tan, A. \mnraso {\bf 200}, 115 (1982)

\pp
Chokshi, A., \& Turner, E. L. \mnraso {\bf 259}, 421 (1992)
\bigskip

\nhi$\bullet$
Dynamical models of galaxies as linear combinations of individual orbits are
discussed in 

\pp
Schwarzschild, M. \apjo {\bf 232}, 236 (1979)

\pp
Richstone, D. O., \& Tremaine, S. \apjo {\bf 327}, 82 (1988)

\pp
van der Marel, R. P., Cretton, N., de Zeeuw, P. T., \& Rix, H.-W. \apjo
      {\bf 493}, 613 (1998)

\pp
Gebhardt, K., \etal \ajo {\bf 119}, 1157 (2000)

\bigskip

\nhi$\bullet$
The BH detection in NGC 3115 is discussed in

\pp
Kormendy, J., \& Richstone, D. \apjo {\bf 393}, 559 (1992)

\pp
Kormendy, J., \etal \aplo {\bf 459}, L57 (1996)
\bigskip

\nhi$\bullet$
The BH detection in M{\ts}31 is discussed in

\pp
Dressler, A., \& Richstone, D. O. \apjo {\bf 324}, 701 (1988)

\pp
Kormendy, J. \apjo {\bf 325}, 128 (1988)
\bigskip

\nhi$\bullet$
Tremaine's model for the double nucleus of M{\ts}31 and new evidence for that 
model are in 

\pp
Tremaine, S. \ajo {\it 110}, 628 (1995)

\pp
Kormendy, J., \& Bender, R. \apjo {\bf 522}, 772 (1999)
\bigskip

\nhi$\bullet$
{\it HST\/} spectroscopy of the double nucleus of M{\ts}31 is presented in

\pp
Statler, T. S., King, I. R., Crane, P., \& Jedrzejewski, R. I. \ajo {\bf 117},
894 (1999)
\bigskip

\nhi$\bullet$
The following are thorough reviews of the Galactic center:

\pp
Genzel, R., Hollenbach, D., \& Townes, C. H. {\it Rep. Prog. Phys.} {\bf 57},
     417 (1994)

\pp
Morris, M., \& Serabyn, E. \annrevo {\bf 34}, 645 (1996)
\bigskip

\nhi$\bullet$
The latest measurement of the size of the Galactic center radio source is by

\pp
Lo, K. Y., Shen, Z.-Q., Zhao, J.-H., \& Ho, P. T. P. \aplo {\bf 508}, L61 (1998)
\bigskip

\nhi$\bullet$
The remarkable proper motion measurements of stars near Sgr A* and resulting
conclusions about the Galactic center BH are presented in

\pp
Genzel, R., Eckart, A., Ott, T., \& Eisenhauer, F. \mnraso {\bf 291}, 219 (1997)

\pp 
Ghez, A. M., Klein, B. L., Morris, M., \& Becklin, E. E. \apjo {\bf 509}, 678 
(1998)
\bigskip

\nhi$\bullet$
Arguments against compact dark star clusters in NGC 4258 and the Galaxy are 
presented in

\pp
Maoz, E. \aplo {\bf 494}, L181 (1998)
\bigskip

\end